\documentclass[aps,prb,twocolumn,showpacs,preprintnumbers,amsmath,amssymb,superscriptaddress]{revtex4-1}

\usepackage{txfonts}
\usepackage{graphicx}
\usepackage{color}
\usepackage{dcolumn}
\usepackage{bm}
\usepackage{feynmp}

\newcommand{\nc}{\newcommand}
\nc{\be}{\begin{eqnarray}}
\nc{\ee}{\end{eqnarray}}
\nc{\bea}{\begin{eqnarray}}
\nc{\eea}{\end{eqnarray}}
\nc{\bean}{\begin{eqnarray*}}
\nc{\eean}{\end{eqnarray*}}
\nc{\mb}{\mbox}
\nc{\rnc}{\renewcommand} 
\nc{\vk}{{\bm k}}
\nc{\vx}{\mb{\bf x}}
\nc{\br}{\mb{\bf r}}
\nc{\bv}{\mb{\bf v}}
\nc{\bp}{\mb{\bf p}}
\nc{\ve}{\mb{\bf e}}
\nc{\vz}{\hat {\mb{\bf z}}}
\nc{\vp}{\mb{\boldmath$p$}}
\nc{\vb}{\mb{\boldmath$b$}}
\nc{\rr}{\mb{\boldmath$r$}}
\nc{\vR}{\mb{\boldmath$R$}}
\nc{\vj}{\mb{\boldmath$j$}}
\nc{\vg}{\mb{\boldmath$g$}}
\nc{\vm}{\mb{\boldmath$m$}}
\nc{\vd}{\mb{\boldmath$d$}}
\nc{\hd}{\mb{\boldmath$\hat{d}$}}
\nc{\vD}{\mb{\boldmath$D$}}
\nc{\vF}{\mb{\boldmath$F$}}
\nc{\vG}{\mb{\boldmath$G$}}
\nc{\vI}{\mb{\boldmath$I$}}
\nc{\vW}{\mb{\boldmath$W$}}
\nc{\x}{\mb{\boldmath$x$}}
\nc{\A}{\mb{\boldmath$A$}}
\nc{\va}{\mb{\boldmath$a$}}
\nc{\vv}{\mb{\boldmath$v$}}
\nc{\vq}{\mb{\boldmath$q$}}
\nc{\vn}{\mb{\boldmath$n$}}
\nc{\vJ}{\mb{\boldmath$J$}}
\nc{\vS}{\mb{\boldmath$S$}}
\nc{\vs}{\mb{\boldmath$\sigma$}}
\nc{\vE}{\mb{\boldmath$E$}}
\nc{\vB}{\mb{\boldmath$B$}}
\nc{\vM}{\mb{\boldmath$M$}}
\nc{\vL}{\mb{\boldmath$L$}}
\nc{\vpsi}{\mb{\boldmath$\psi$}}
\nc{\vphi}{\mb{\boldmath$\varphi$}}
\nc{\Vphi}{\mb{\boldmath$\phi$}}
\nc{\Vomega}{\mb{\boldmath$\Omega$}}
\nc{\ipsi}{\it{\Psi}}
\nc{\vepsilon}{\mb{\boldmath$\epsilon$}}
\nc{\valpha}{\mb{\boldmath$\alpha$}}
\nc{\vgamma}{\mb{\boldmath$\gamma$}}
\nc{\vomega}{\mb{\boldmath$\omega$}}
\nc{\vmu}{\mb{\boldmath$\mu$}}
\nc{\vt}{\mb{\boldmath$\tau$}}
\nc{\vT}{\mb{\boldmath$T$}}
\nc{\vpi}{\mb{\boldmath$\pi$}}
\nc{\nab}{\bm{\nabla}}
\nc{\ov}{\overline}
\nc{\cdott}{\!\cdot\!}
\nc{\cdottt}{\!\!\cdot\!}
\nc{\LL}{\Big{\langle}}
\nc{\RR}{\Big{\rangle}}
\nc{\LR}{\Bigm{|}}
\nc{\vP}{\mb{\boldmath$P$}}
\nc{\nnn}{\nonumber\\}
\nc{\ltsim}{\protect\raisebox{-0.5ex}{$\:\stackrel{\textstyle <}{\sim}\:$}}
\nc{\gtsim}{\protect\raisebox{-0.5ex}{$\:\stackrel{\textstyle >}{\sim}\:$}} 
\nc{\ltsimscript}{\protect\raisebox{-0.5ex}{$\stackrel{\scriptstyle <}{\sim}$}} 
\nc{\gtsimscript}{\protect\raisebox{-0.5ex}{$\stackrel{\scriptstyle >}{\sim}$}} 

\rnc{\figurename}{FIG.}
\def\slash#1{\!\not\!#1}

\nc{\psibar}{\overline{\psi}}
\nc{\cbar}{\overline{c}}
\nc{\intx}{\int d^4x}
\nc{\inty}{\int d^4y}
\nc{\intk}{\int \frac{d^4k}{(2\pi)^4}}
\nc{\Mhat}{\hat{\bm M}}

\begin{document}

\title{
Microscopic theory of electrically induced spin torques in magnetic Weyl semimetals
}

\author{Daichi Kurebayashi}
\email{daichi.kure@imr.tohoku.ac.jp}
\affiliation{
Institute for Materials Research, Tohoku University, Sendai 980-8577, Japan
}
\author{Kentaro Nomura}
\affiliation{
Institute for Materials Research, Tohoku University, Sendai 980-8577, Japan
}

\date{\today}

\begin{abstract}
We theoretically study electrical responses of magnetization in Weyl semimetals.
The Weyl semimetal is a new class of topological semimetals, possessing hedgehog type spin textures in momentum space.
Because of this peculiar spin texture, an interplay of electron transport and spin dynamics might provide new method to electrical control of magnetization.
In this paper, we consider the magnetically doped Weyl semimetals, and systematically study current- and charge-induced spin torque exerted on the local magnetization in three-dimensional Dirac-Weyl metals.
We determine all current-induced spin torques including spin-orbit torque, spin-transfer torque, and the so-called $\beta$-term, up to first order with respect to spatial and temporal derivation and electrical currents.
We find that spin-transfer torque and $\beta$-term are absent while spin-orbit torque is proportional to the axial current density.
We also calculate the charge-induced spin torque microscopically.
We find the charge-induced spin torque originates from the chiral anomaly due to the correspondence between spin operators and axial current operators in our model.

\end{abstract}

\maketitle

\section{INTRODUCTION}
Electrical manipulation of magnetization is of major interest in the field of spintronics for achieving low-energy consumption electronic devices \cite{Zutic2004}.
In ferromagnetic metals, spin-polarized currents are widely used to control magnetization dynamics such as spin-transfer torque and spin pumping \cite{Slonczewski1996,Berger1996,Ralph2008,Brataas2012}.
Utilizing charge currents, however, suffers from the Joule heating, limiting the energy efficiency.
Recently, the application of the topological properties of materials has drawn much interest to achieve a more efficient manipulation of the magnetization.
For instance, in magnetically doped topological insulators and ferromagnetic insulators deposited on the surface of a topological insulator\cite{Hasan2010,Qi2011}, electrical manipulation of magnetic textures \cite{Nomura2010,Hurst2015}, electric field- or current-induced magnetization switching\cite{Garate2010,Yokoyama2010,Sakai2014,Fan2014}, and spin-charge conversion\cite{Shiomi2014,Kondou2016,Okuma2016} have been considered theoretically and experimentally.
Topological materials are often realized by strong spin-orbit coupling, thus the spin degrees of freedom couple to the momenta, known as the spin-momentum locking.
Spin-momentum locking may provide a new means to achieve a reliable and less dissipative control of magnetization.

As a new class of topological materials, Weyl semimetals are intensively researched.
Weyl semimetals are characterized by pairs of bulk gapless points each distinguished by their chirality.\cite{Nielsen1981,Hosur2013}
Close to the gapless points, excitations are described by a three-dimensional linear dispersion which is analogous to the Weyl fermion in high-energy physics.
To realize Weyl semimetals, at least time-reversal or inversion symmetry must be broken.\cite{Zyuzin2012,ShuichiMurakami2007}
Weyl semimetals whose time-reversal symmetry is broken by the magnetic ordering of local moments are especially promising for spintronics applications, because they possess both magnetic and topological properties.
To realize magnetic Weyl semimetals, there are a lot of theoretical proposals such as pyrochlore iridates\cite{Wan2011}, multilayers of topological insulators and normal insulators\cite{Burkov2011}, and magnetically doped topological insulators\cite{Sekine2013,Bulmash2014,Kurebayashi2014}
Recently, the noncollinear antiferromagnets $\rm Mn_3X$ ($\rm X=Sn,Ge$) were theoretically and experimentally proposed as time-reversal broken Weyl semimetals.

In our previous papers\cite{Nomura2015,Kurebayashi2016}, we found that spin torque can be generated by an applied gate voltage in a magnetic field, so-called charge-induced spin torque, and proposed a spintronic device incorporating magnetic Weyl semimetals.
Our derivation was phenomenological, based on the peculiar coupling between magnetization and charge density emerging from the topological nature of Weyl semimetals.
The microscopic origin of the effect still has to be clarified.
One of the main focuses of this paper is to give the physical understanding of the charge-induced spin torque.
In addition to the charge-induced spin torque, the magnetization responses to the applied currents are also not well known.
In this paper, we study magnetization dynamics in magnetic Weyl semimetals by considering spin torques.
We present a microscopic calculation of spin torques, specifically spin-orbit torque, spin-transfer torque, $\beta$-term, and the charge-induced torque in our continuum model for magnetic Weyl semimetals.
Consequently, we obtain the analytical form of the spin-orbit torque and the charge-induced spin torque in magnetic Weyl semimetals.
The results concerning the charge-induced spin torque is consistent with that proposed in Ref. 26, and is understood by the chiral anomaly.
We also find that the spin-transfer torque and $\beta$-term are absent in magnetic Weyl semimetals.
The absence of spin-transfer torque and $\beta$-term is understood by the correspondence of transport and spin phenomena which our model possesses.

\section{FORMALISM}
Magnetization dynamics is described by the Landau-Lifshitz-Gilbert equation as
\bea
\frac{d \hat{\bm M}}{d t}=\gamma_0\bm H\times \hat{\bm M}+\alpha_0 \hat{\bm M}\times\frac{d \hat{\bm M}}{d t} +\bm T_e,
\eea
where $\gamma_0$ is the gyromagnetic ratio, $\bm H$ is an external magnetic field, and $\alpha_0$ is the damping constant.
The effect of the background of conduction electrons is described in terms of spin torques as
\bea
\bm T_e(\bm r)=JS\hat{\bm M}\times \langle\hat{\bm \sigma}(\bm r)\rangle
\eea
where $\langle\hat{\bm \sigma}(\bm r)\rangle$ is the spin density of conduction electrons.\cite{Tserkovnyak2006}
The spin torques can be obtained by calculating $\langle\hat{\bm \sigma}(\bm r)\rangle$.
In a weak spin-orbit coupled system, the spin density can be expanded as
\bea
\nonumber \langle\hat{\bm \sigma}(\bm r)\rangle=a_0\frac{d \hat{\bm M}}{d t}+b_0\hat{\bm M}\times\frac{d \hat{\bm M}}{d t}+(\bm a\cdot \bm \nabla) \hat{\bm M}+\hat{\bm M}\times(\bm b\cdot \bm \nabla) \hat{\bm M}
\label{eq_st}\\
\eea
in the first order of time derivative and spatial gradient.
The first two terms are a correction to the Gilbert damping and the spin Berry phase term 
renormalizing the magnitude of spin, respectively.
The last two terms describe the current-induced spin torques, the spin-transfer torque and so-called $\beta$-term.
With conventional Schr\"{o}dinger like electrons, the spin-transfer torque and the $\beta$-term are proportional to a spin current $\bm j_S$, described by $\bm T_{\rm STT}\propto (\bm j_S\cdot \bm \nabla)\hat{\bm M}$ and $\bm T_{\beta}\propto \Mhat\times (\bm j_S\cdot \bm \nabla)\Mhat$, respectively.\cite{Kohno2006}
In strong spin-orbit coupled systems, other types of current-induced spin torques are expected.
For example, in two-dimensional Rashba systems or surface of the three-dimensional topological insulators, spin-orbit torque is expected due to spin-momentum locking, $\bm T_{\rm SO}\propto \hat{\bm z}\times \bm j$ where $\hat{\bm z}$ is the normal vector perpendicular to the surface and $\bm j$ is an electric current.\cite{Yokoyama2010,Sakai2014}

In addition to current-induced torques, the spin torque generated by a local charge density in the presence of an external magnetic field is expected in anomalous Hall ferromagnets.\cite{Nomura2015}
There is a relationship, in anomalous Hall ferromagnets, between the charge density and the magnetization, $\rho_{\rm ind}\propto \sigma_{\rm AHE}\Mhat\cdot\bm B$\cite{Xiao2010,Zyuzin2012a,Nomura2015} where $\rho_{\rm ind}$ is the induced charge-density and $\sigma_{\rm AHE}$ is the magnitude of the anomalous Hall conductivity.
The relation states that the local charge density increases when the magnetization is parallel to the external magnetic field and decreases when it's antiparallel.
As an inverse effect of this relation, a spin torque is generated by modifying the local charge density to flip the magnetization.
This spin torque is called charge-induced spin torque.
Time-reversal symmetry broken Weyl semimetals are known to host the anomalous Hall effect\cite{Yang2011}, so charge-induced spin torque is also to be expected.

In this paper, we calculate the non-equilibrium electronic spin polarization in magnetic Weyl semimetals generated by electric voltage and currents.
As a simple model, we consider a Weyl semimetal realized in a Dirac semimetal coupled to the local moments of magnetic dopants.\cite{Bulmash2014,Kurebayashi2014}
This might be realized in magnetically doped topological insulator systems such as chromium-doped $\rm Bi_2Se_3$ and $(\rm Bi, Sb)_2Te_3$, where increasing chromium concentration reduces the strength of spin-orbit coupling and leads to a topological phase transition into a normal insulator phase with band gap closing.\cite{Jin2011,Kim2013}
For clarify,  we note that the Dirac semimetal considered here is not a symmetry-protected Dirac semimetal such as $\rm Cd_2As_3$ and $\rm Na_3Bi$.\cite{Wang2012,Neupane2014,Liu2014}
The low-energy effective Hamiltonian of the magnetic Weyl semimetals\cite{Burkov2011,Zyuzin2012a} is
\bea
\nonumber H_{\rm WSM}&=&\int d\bm r \psi^\dagger(\bm r) \biggl[ \left\{v_F\tau_z \bm \sigma \cdot\left(-i\bm \nabla + e \bm A(\bm r,t)\right)-E_F\right\}-\tau_z \Delta \\
&&\hspace{2.5cm} +JS\hat{\bm M}(\bm r,t)\cdot \bm \sigma \biggl]\psi(\bm r)+\hat{V}_{\rm imp}.\label{H_0}
\eea
The first term describes a Dirac semimetal consisting of degenerate three-dimensional massless electrons where $v_F$ is the Fermi velocity, $\bm \sigma=(\sigma_x,\sigma_y,\sigma_z)$ are Pauli matrices representing real spin operators, and $\tau_z=\pm1$ labels the chirality of two Weyl nodes.
In the second and third terms, $\Delta$ parameterizes the inversion-symmetry breaking, being finite in noncentrosymmetric materials, and $J$ is the exchange coupling constant between conduction electron spins and local magnetic moments, $S$ is spin of localized magnetic moments.
$\hat{\bm M}=(M_x,M_y,M_z)$ is a normalized directional vector of the magnetic moments.
The last term is the coupling to random impurities given as $\hat{V}_{\rm imp}=\int d\bm r \psi^\dagger(\bm r)V_{\rm imp}(\bm r)\psi(\bm r)$.
We consider non-magnetic impurities
$
V_{\rm imp}(\bm r)=\sum_i u(\bm r-\bm R_i)
$
with short-range potential $u(\bm r)\rightarrow u_0 \delta (\bm r)$.
A Gaussian average is taken for impurity positions as $\overline{V_{\rm imp}(\bm r)V_{\rm imp}(\bm r')}=n_iu_0^2\delta(\bm r-\bm r')$ where $n_i$ is the concentration of impurities.
Without impurities $(\hat{V}_{\rm imp}=0)$, the band structure of this Hamiltonian consists of two Weyl nodes with separation $2JS\hat{\bm M}/v_F$ in momentum space and $2\Delta$ in energy.
For clarification, $\Delta$ is different from the chiral chemical potential describing Fermi energy difference between two Weyl nodes.

The Green's function of the Hamiltonian Eq. (\ref{H_0}) is given as
\bea
G_\lambda (i\omega_n,\bm k)=\left[i\omega_n+E_F+\lambda \Delta-\lambda v_F\bm k\cdot \bm \sigma -\Sigma_\lambda(i\omega_n,\bm k)\right]^{-1}
\eea
where $\Sigma_\lambda$ is the self energy induced by impurity scattering, and $\lambda=\pm1$ are the eigenvalues of $\tau_z$ labeling the chirality.
In the first Born approximation, the imaginary part of the self energy is given as
$
{\rm Im}\ \Sigma_\lambda=n_iu_0^2 \sum_{\bm q} {\rm Im}\ G_\lambda(i\omega_n,\bm q)=\pi n_iu_0^2D_\lambda(E_F){\rm sgn}(\omega_n)
\equiv \eta_\lambda{\rm sgn}(\omega_n)
$
where $D_\lambda(E_F)$ is the density of states at the Fermi energy for a node with the chirality $\lambda$, and $\eta_\lambda$ is the electron damping rate by impurity scattering.
We neglect the real part of the self energy because they can be absorbed into the Fermi energy.
In the following, we assume that the damping rate is the same in both nodes, $\eta\equiv\eta_+=\eta_-$, and $\eta<<E_F$ corresponding to a weak impurity scattering regime.
We calculate spin torques in the lowest order of $\eta$.

\section{Current-induced spin torques}
\begin{figure}[tbp]
\begin{center}
\includegraphics[width=0.9\linewidth]{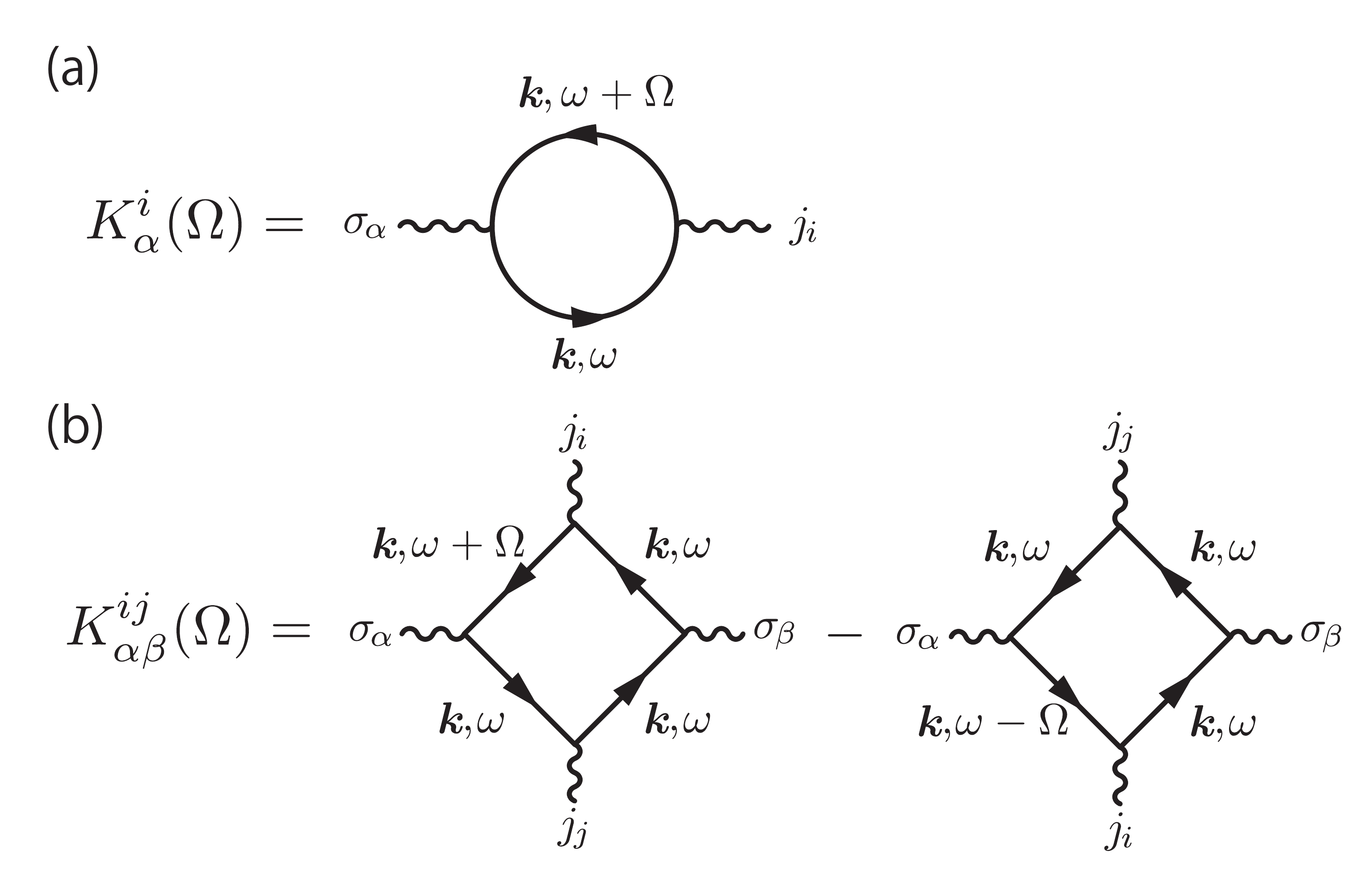}
\caption{
Feynman diagrams for the current-induced spin torques, (a) the spin-orbit torque and (b) the spin-transfer torque and the $\beta$-term.
}
\label{fig3}
\end{center}
\end{figure}

In this section, we discuss the current-induced magnetization dynamics.
For current-induced spin torques, we consider small transverse magnetization fluctuations, $\delta\bm M=(\delta M_x,\delta M_y,0), \|\delta\bm M\|<<1$, in the presence of the static electronic fields.
Coupling to an electric field is introduced by $H'=\int d\bm r \bm j(\bm r)\cdot \bm A(\bm r)$ where $\bm j(\bm r)=\psi^\dagger(\bm r)\hat{\bm j}\psi(\bm r)$ is a current operator.
Here the saturation magnetization axis is taken in $z$-axis.
We calculate the conduction electron's spin density in response to an electric field within linear response theory as
\bea
\nonumber \langle \sigma_\alpha(\bm q)\rangle&=&\chi_\alpha^i (\bm q) E_i,\\
\chi_{\alpha}^i(\bm q) &=& \lim_{\Omega\rightarrow 0}\frac{K_\alpha^i(\bm q,\Omega+i0)-K_\alpha^i(\bm q, 0)}{i\Omega}
\label{spin_current}
\eea
where
$
K_\alpha^i(\bm q,\Omega)=i \int^\infty_0 dt e^{i\Omega t}\langle[\sigma_\alpha(\bm q,t),j_i]\rangle
$
is a dynamical correlation function between the electron spin and the electric currents.
Here we consider a uniform electric field, thus the wave vector $\bm q$ comes only from the magnetization fluctuations.
Furthermore, we assume that the spatial variation of the magnetization fluctuation is small and expand the response function up to linear order in $\bm q$ and $\delta \bm M$ as
\bea
K_\alpha^i(\bm q,\Omega)\approx K_\alpha^i(\Omega) + 2K_{\alpha \beta}^{ij}(\Omega) q_j \delta M_{\beta},
\label{res_func}
\eea
where indices $\alpha,\beta=1,2$ denote direction in spin space and $i,j=1\sim3$ denote direction in momentum space.
The first term corresponds to the spin-orbit torque while the second term corresponds to torque appearing when the magnetization is not uniform such as the spin-transfer torque and the $\beta$-term.
The response functions are obtained by
\bea
K_\alpha^i(i\Omega_m)&=&k_BTe\sum_{n,\bm k,\lambda}{\rm Tr}\left[\sigma_\alpha G_{\lambda}v_{i,\lambda}G^+_\lambda \right]
\label{res_sot}\\
\nonumber K_{\alpha \beta}^{ij}(i\Omega_m)&=&k_BTeJS\sum_{n,\bm k,\lambda }{\rm Tr}\left[\sigma_\alpha G_\lambda v_{\lambda,j}G_\lambda\sigma_\beta G_\lambda v_{\lambda,i}  G^+_\lambda\right.\\
&&\left. -\sigma_\alpha G^-_\lambda v_{\lambda,i}G_\lambda\sigma_\beta G_\lambda v_{\lambda,j}  G_\lambda\right]
\label{res_current}
\eea
where $\bm v_{\lambda}$ is velocity operators of electrons with the chirality $\lambda$, and $G_\lambda, G_\lambda^\pm$ denote $G_\lambda(i\omega_n,\bm k), G_\lambda(i\omega_n\pm i\Omega_m,\bm k)$, respectively, and are expressed as Feynman diagrams in Fig.\ref{fig3}.

Before moving on to the detail calculation of the response functions, let us examine $K_{\alpha\beta}^{ij}$ with symmetry arguments.
After some calculation shown in Appendix A, the response function can be written as
\bea
K_{\alpha\beta}^{ij}=A\delta_{\alpha\beta}\delta_{ij}+B\delta_{\alpha i}\delta_{\beta j}+C\delta_{\alpha j}\delta_{\beta i}.
\label{k_current}
\eea
By substituting this into Eq.(\ref{spin_current}) and (\ref{res_func}), the spin density associated with the electric field and magnetic textures is 
\bea
\langle\bm \sigma \rangle= A(\bm E\cdot\bm \nabla)\delta\bm M+B\bm E(\bm \nabla\cdot\delta\bm M)+C\bm \nabla(\bm E\cdot \delta\bm M).
\label{eq_current}
\eea
In the linearized Hamiltonian, Eq.(\ref{H_0}), there is a correspondence between the axial velocity operators and the spin operators, i.e.
\bea
\bm v_5=\tau_z\frac{\partial H_{\rm WSM}}{\partial \bm k}= v_F\psi^\dagger\bm \sigma \psi.
\eea
This suggests that the spin operators and the axial current operators are identical except for a constant factor\cite{Taguchi2015}
\bea
\bm j_5=-e\psi^\dagger \bm v_5 \psi=-ev_F\psi^\dagger \bm \sigma \psi.
\label{correspondence}
\eea
The electron spins couple to the magnetization by the exchange interaction as $-\bm M\cdot \bm \sigma$, while the axial currents couple to the axial vector potential as $- \bm A_5\cdot\bm j_5$.
Due to the correspondence between the spin operators and the axial current operators, magnetization and the axial vector potentials couple to electrons in the same way, namely the electrons cannot distinguish the perturbations.
By using this correspondence, Eq.(\ref{eq_current}) can be interpreted in the axial transport picture as
\bea
\langle\bm j_5\rangle =A(\bm E\cdot\bm \nabla)\bm A_5+B\bm E(\bm \nabla\cdot\bm A_5)+C\bm \nabla(\bm E\cdot \bm A_5)
\eea
in the expression explicitly depending on the axial vector potentials.
This expression is not gauge invariant under the local chiral gauge transformation.
Namely, these terms are prohibited when local chiral gauge symmetry is present.
Consequently, coefficients of these terms including the spin-transfer torque and the $\beta$-term are going to be zero.\cite{note1}
On the other hand, the spin density associated with $K_\alpha^i$ is gauge invariant since it only includes electric field.
Therefore only the spin susceptibility of spin-orbit torque is finite while the others must be zero.
A similar discussion exists for a ferromagnetic insulator on the surface of a topological insulator \cite{Sakai2014}, however the responsible symmetries are different, i.e. the U(1) gauge symmetry for the TI surface but the chiral gauge symmetry for the Weyl semimetals.
Note that the correspondence between the spin operators and the axial velocity operators retains when the band dispersion close to the Fermi energy is described by $k$-linear term.
If the dispersion deviates from linear, including $k^2$-term, there is no restriction from local chiral gauge symmetry, and the spin-transfer torque and the $\beta$-term are also expected.

The leading order of the spin response function associated with the spin-orbit torque in damping rate, $\eta$, is calculated in Appendix A, giving
\bea
\nonumber K^i_\alpha(\Omega)&=&i\Omega\delta_{\alpha i}\frac{e}{6\pi^2v_F}\sum_\lambda \frac{\left(E_F+\lambda \Delta\right)^2}{\eta}\\
&=&i\Omega\delta_{\alpha i}\frac{eE_F\Delta}{3\pi^2v_F^2\eta},
\label{k_sot}
\eea
and the spin density induced by the applied electric fields is given by
\bea
\langle\bm \sigma\rangle=\frac{eE_F\Delta}{3\pi^2v_F^2\eta}\bm E.
\label{result_sot}
\eea
The associated spin torque is therefore
\bea
\bm T_e(\bm r)=\frac{eJSE_F\Delta}{3\pi^2v_F^2\eta} \hat{\bm M}\times\bm E=-\frac{JS}{ev_F}\hat{\bm M}\times \langle\bm j_5\rangle
\eea
where $\langle \bm j_5\rangle=-(e^2E_F\Delta/3\pi^2v_F\eta)\bm E$ is the axial current density.
The second equality is obtained from the correspondence between the spin and the axial transport phenomena.
Equation (\ref{result_sot}) indicates that the spin density is induced in the direction of the applied electric field.
The effect can be understood by the Rashba-Edelstein effect.
In the Weyl Hamiltonian, momentum and spin are fixed to be parallel or antiparallel depending on the chirality, the spin-momentum locking.
Suppose that the electric field is applied in the $+x$ direction, the Fermi surface is shifted along $-k_x$ direction, and the number of the electrons with $+k_x$ decreases whereas the number with $-k_x$ increases.
Because of spin-momentum locking, the imbalance of the momentum induces positive $x$ component of the net spin density for the right-handed and negative for the left-handed.
When $\Delta=0$, meaning two Weyl nodes are located at the same energy, the induced spin density of each node has same magnitude but opposite direction, thus the contributions cancel out.
Once $\Delta$ becomes non-zero, the cancellation is incomplete and the net spin density becomes finite.
In the presence of inversion-symmetry, $\Delta$ has to vanish.
Therefore, to observe the spin-orbit torque, both the inversion and time-reversal symmetries must be broken.

\section{Charge-induced spin torque}
In this section, we microscopically derive the charge-induced spin torque, and discuss the relation to the chiral anomaly.
The charge-induced spin torque is described as a spin response to the external electric voltage and magnetic fields.
Coupling to external fields is introduced by $H'=\int d\bm r \bm j(\bm r)\cdot \bm A(\bm r)$ where $\bm j(\bm r)=\psi^\dagger(\bm r)\hat{\bm j}\psi(\bm r)$ and $H''=e\int d\bm r \psi^\dagger(\bm r)\phi(\bm r)\psi(\bm r)$ where $\phi(\bm r)$ is an electric potential.
We calculate the electron's spin density $\langle\bm \sigma\rangle$ up to linear order in both the electric scalar potential, $\phi$, and the vector potential $\bm A$.
The electron's spin density can be obtained as
\bea
\langle\sigma_\alpha(\bm q)\rangle=2 \chi_\alpha^i (\bm q) \phi A_i(\bm q)
\label{eq_cit}
\eea
where $\alpha$ and $i$ denote spin and velocity direction, respectively, and $\chi_\alpha^i (\bm q)$ is the response function of the spin polarization given by
\bea
\nonumber \chi_\alpha^i (\bm q)& = &-e^2v_F\sum_{\omega,\bm k,\lambda} \left\{ {\rm Tr}\left[\sigma_\alpha \tilde{G}_\lambda^- G_\lambda v_{i,\lambda}G_\lambda \right]\right .\\
&&\left .+{\rm Tr}\left[\sigma_\alpha G_\lambda v_{i,\lambda}G_\lambda \tilde{G}_\lambda^+\right]\right\},
\label{cit_chi}
\eea
where $\tilde{G}_\lambda^\pm \equiv G_\lambda(\omega,\bm k\pm\bm q)$ and $G_\lambda\equiv G_\lambda(\omega,\bm k)$.
The response function is represented diagrammatically in Fig. \ref{FD_CIT}.
This is calculated in Appendix B, and the result up to $\mathcal{O}(\eta^0)$, the leading order in the scattering rate for the charge-induced spin torque, is given by
\bea
\chi_\alpha^i(\bm q)=-i\frac{e^2}{4\pi^2v_F}\epsilon_{\alpha i j}q_j,
\label{cit_chi2}
\eea
where $\epsilon_{ijk}$ is antisymmetric under any permutation of indexes.
From Eq.(\ref{eq_cit}), the spin density induced by the external fields is given as
\bea
\langle \sigma_\alpha(\bm q)\rangle=\frac{e^2}{2\pi^2v_F}\phi\epsilon_{\alpha i j} iq_iA_j(\bm q)=\frac{e^2}{2\pi^2v_F}\phi B_\alpha(\bm q),
\label{spin_cit}
\eea
where $B_\alpha(\bm q)=\epsilon_{\alpha i j}iq_iA_j(\bm q)$ is the external magnetic field.
Therefore, the spin torque associated with this process is expressed by
\bea
\bm T_e(\bm r)=\frac{e^2JS}{2\pi^2v_F}\phi\hat{\bm M}\times\bm B(\bm r)=\sigma_{\rm AHE}\phi \hat{\bm M}\times\bm B(\bm r),
\label{cit_result}
\eea
where $\sigma_{\rm AHE}=\frac{e^2SJ}{2\pi^2v_F}$ is the anomalous Hall conductivity of magnetic Weyl semimetals.
By replacing the electric potential with the shift in the chemical potential, $\phi\rightarrow -\delta \mu_F/e$, Eq. (\ref{cit_result}) is consistent with our previous result obtained phenomenologically. \cite{Nomura2015}

\begin{figure}[tbp]
\includegraphics[width=1\linewidth]{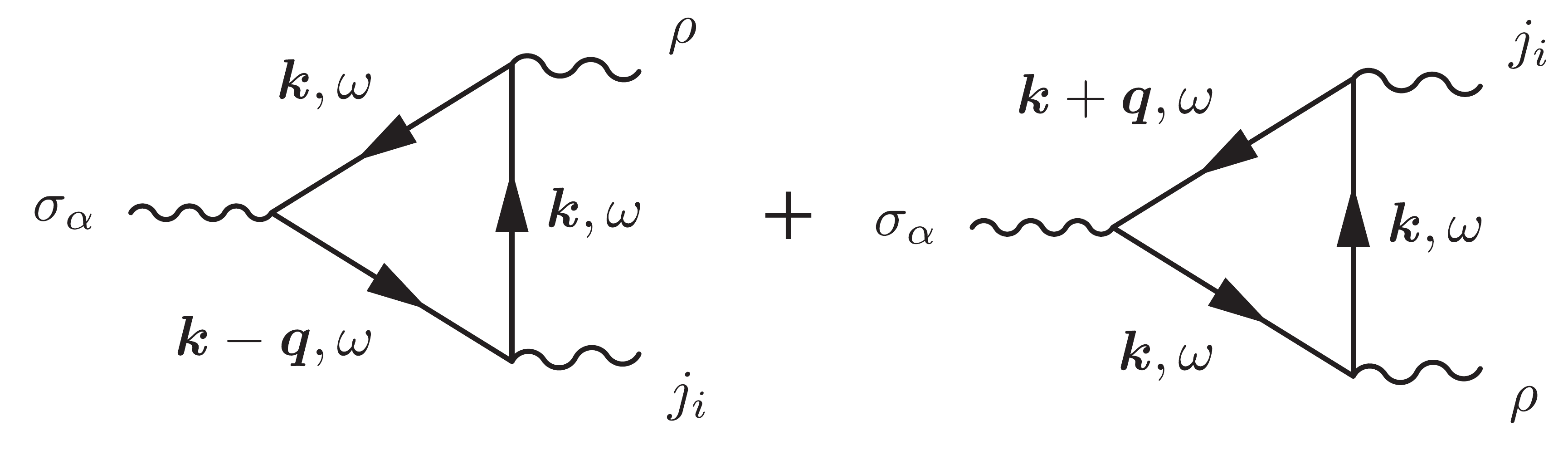}
\caption{Feynman diagrams for the charge-induced spin torque.}
\label{FD_CIT}
\end{figure}

The physical origin of the spin polarization, Eq.(\ref{spin_cit}), can be understood by considering the Landu levels of chiral fermions.
Here we consider Weyl fermions in a uniform magnetic field applied along the $z$ direction.
Defining canonical momenta as $\pi_i=-i\nabla_i+e A_i+\tau_z(JS/v_F)M_i$ where $i=x,y$, the Hamiltonian given in Eq.(\ref{H_0}) can be written as\cite{Zyuzin2012a}
\bea
\hspace{-5mm}\mathcal{H}_{\rm WSM}=\tau_zv_F\left(k_z+\tau_z\frac{JS}{v_F}M_z\right)\sigma_z+\tau_zv_F\left(\pi_x\sigma_x+\pi_y\sigma_y\right).
\label{H_LL}
\eea
The spin operators are described by
\bea
\sigma_i=\tau_z\frac{\delta H_{\rm WSM}}{v_F\delta \pi_i},\ 
\sigma_z=\tau_z\frac{\delta H_{\rm WSM}}{v_F\delta k_z}.
\eea
The correspondence between the spin operators and the axial velocity operators still holds in the presence of magnetic fields.
The Hamiltonian, Eq.(\ref{H_LL}) can be solved obtaining the energy eigenvalues 
\bea
E_0^\lambda(k_z)&=&-\lambda v_F k_z-JSM_z,\\
E_n^\lambda(k_z)&=&-v_F {\rm sgn}(n)\sqrt{\left(k_z+\lambda \frac{JSM_z}{v_F}\right)^2+2eB_z|n|}
\label{eq_LL}
\eea
where $n=\pm 1,\pm2 ,\cdots$ are the Landau indices, $\lambda=\pm 1$ labels the chirality.
The $z$ component of the total spin density is given by the sum of contributions from all states below the Fermi energy
\bea
\langle\sigma_z\rangle_{\rm tot}=\frac{1}{v_F}\sum_{k_z}\sum_{n}^{E_F>E_n(k_z)}\left[v_{n,k_z}^+-v_{n,k_z}^-\right],
\eea
where $v^\lambda_{n,k_z}=\frac{dE_n^\lambda(k_z)}{dk_z}$ is the group velocity, while $x$ and $y$ components become zero because the Landau levels disperse only along the $k_z$ axis.

Let us first consider contributions from the zeroth Landau levels.
The group velocity of the zeroth Landau levels is $-v_F$ for the right-handed electrons ($\lambda=+1$), and $+v_F$ for the left-handed electrons ($\lambda=-1$).
Therefore, in the zeroth Landu levels, both right and left-handed electrons have down spins with respect to the magnetic field.
By contrast, the non-zero Landau levels do not contribute to the total spin density.
Suppose $\varepsilon$ is a some energy which cuts the non-zero Landau levels below Fermi energy, the momenta associated with $\varepsilon$ are obtained as
\bea
k_\pm^\lambda=-\lambda \frac{JSM_z}{v_F}\pm\sqrt{\left(\varepsilon/v_F\right)^2-2eB_z|n|}.
\eea
The group velocity of the non-zero Landau levels at energy $\varepsilon$ is
\bea
\nonumber v^\lambda_{n,k_\pm^\lambda}=\frac{1}{v_F}\left.\frac{dE_n^\lambda(k_z)}{dk_z}\right|_{k_z=k_\pm^\lambda}
=\mp v_F {\rm sgn}(n)\frac{\sqrt{\left(\varepsilon/v_F\right)^2-2eB_z|n|}}{\varepsilon/v_F}
\eea
when $|\varepsilon|>\sqrt{2eB_z|n|}v_F$.
So there always exist two states possessing opposite sign of the velocity within a Landau level with the chirality $\lambda$.
According to the above arguments, contributions to the spin density from the non-zero Landau levels exactly cancel out.

The total spin density, therefore, is given only by the zeroth Landau levels and estimated as
$
\langle \sigma_z\rangle_{\rm tot}=- n_0(E_F)
$
where $n_0(E_F)$ is electron density in the zeroth Landau levels below the Fermi energy.
When the Fermi energy is shifted by the applied gate voltage, $\phi$, the induced spin density is given by 
\bea
\nonumber \langle \sigma_z \rangle&=&-g_{LL}\int^{E_F-e\phi}_{E_F}dE D(E)\\
&=&\frac{e^2}{4\pi^2v_F}\phi B_z
\label{cit_ll}
\eea
where $g_{LL}=\left|\frac{e\bm B_z}{2\pi}\right|$ is the zeroth Landau level degeneracy and $D(E)=2/2\pi v_F$ is the density of the states for the one-dimensional linear dispersion.
This result is identical to Eq. (\ref{spin_cit}).

Above results are derived from the continuum model where the energy dispersion is linearized and each chirality is separated.
It is, however, not obvious that analysis with the continuum model always gives the same results as a lattice model where the energy-momentum cutoff is naturally introduced and the chirality is no longer a good quantum number.
For example, the chiral magnetic effect in equilibrium, predicted using a continuum model, is actually absent in lattice systems.\cite{Vazifeh2013}
In the following, we numerically examine the spin density of Weyl semimetals on a lattice to check whether the predicted phenomenon survives.

\begin{figure}[tbp]
\begin{center}
\includegraphics[width=1\linewidth]{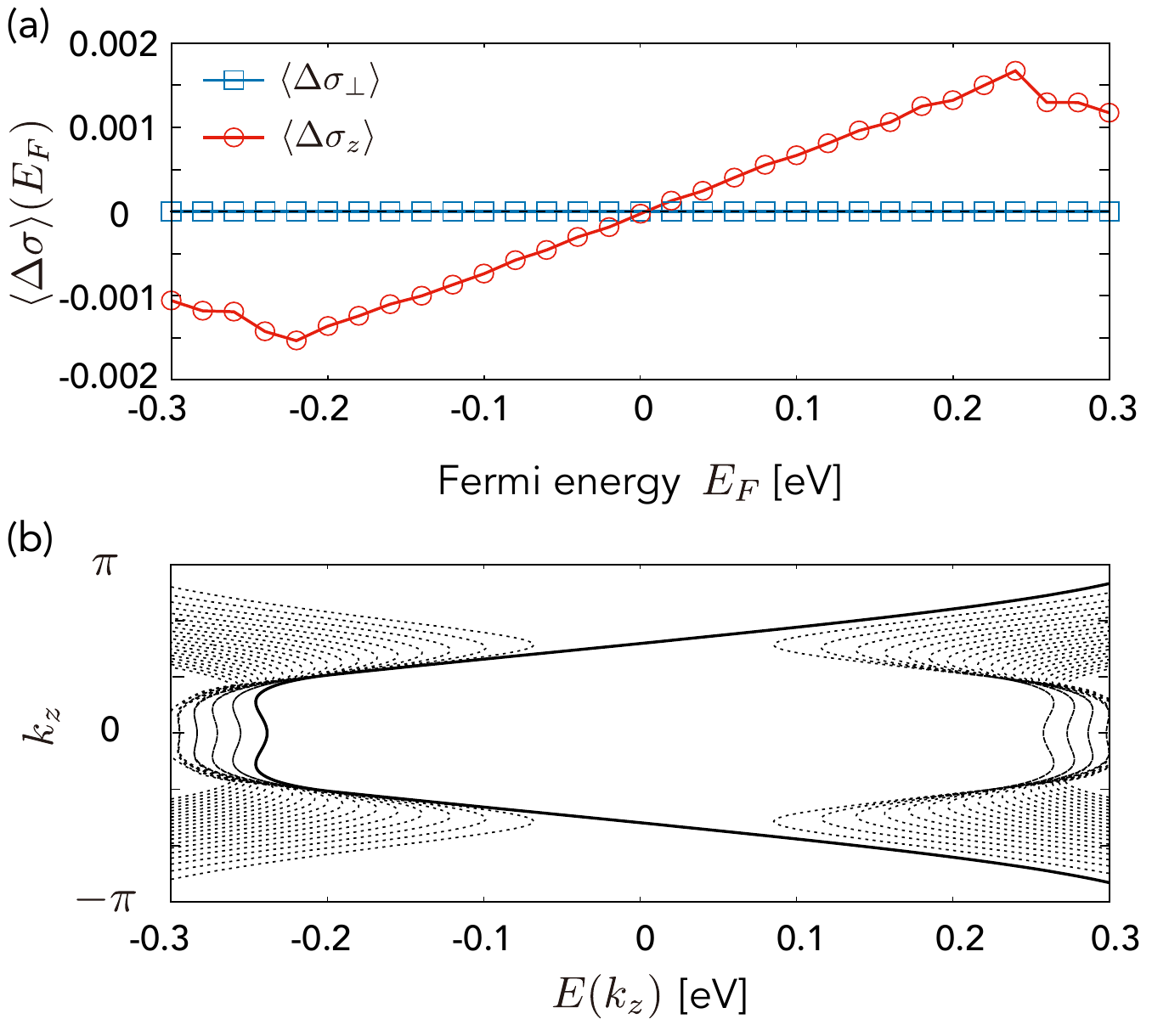}
\caption{(Color online)
(a) Electron spin density calculated as a function of the Fermi energy.
(b) Landau level structure of the lattice Hamiltonian of the Weyl semimetal. The zeroth landau levels are shown as a solid line, while non-zero Landu levels are dashed lines.
The parameters are chosen as $t=-0.248$ eV, $r=0.414$ eV, $J=0.2$ eV, $\hat{\bm M}=\hat{\bm z}$, $\hat{\bm B}=B_0\hat{\bm z}$, and $B_0=\phi_0/128$ where $\phi_0$ is the magnetic flux quantum.
}
\label{fig2}
\end{center}
\end{figure}

We describe a Dirac-Weyl semimetal in a magnetic field using the lattice Hamiltonian\cite{Qi2008}
\bea
H_{\rm Lattice}=\sum_{j=x,y,z}\left(T_j+T_j^\dagger\right)+H_W+H_{\rm exc}
\eea
where $T_j=\sum_{\bm R} c^\dagger_{\bm R+\bm e_j}\left(-it\alpha_j+r\alpha_4\right)e^{iA_j(\bm R)}c_{\bm R}$ are the the translation operators in $j=x,y,z$ directions, $\alpha_j=\sigma_j\tau_z$ and $\alpha_4=-\sigma_0\tau_x$ being $4\times4$ Dirac matrices satisfying the anticommutation relation, $H_W=\sum c^\dagger_{\bm R}(3r\alpha_4) c_{\bm R}$ is the Wilson term in the massless case, and $H_{\rm exc}$ is the exchange Hamiltonian.
The effect of the magnetic field is introduced by the Peierls phase.
We use typical materials parameters for $\rm Bi_2Se_3$ determined to fit {\it ab} {\it initio} calculations.\cite{Zhang2009,Liu2010}
The external magnetic field is taken to be uniform and pointing in the $+z$ direction with the Landau gauge, $\bm A(\bm R)=(0,B_0x,0)$.

We calculate the electron spin density below the Fermi energy, $\langle \sigma_z\rangle(E_F)=\langle c^\dagger \hat{\sigma_z}c \rangle$, by numerical diagonalization of the Hamiltonian.
To measure the response to external fields, we evaluate $\langle\Delta\sigma_z\rangle(E_F)\equiv \left.\langle \sigma_z\rangle(E_F)\right|_{B_z=B_0}- \left.\langle\sigma_z\rangle(E_F)\right|_{B_z=0}$ which is shown in Fig. \ref{fig2} (a). 
When the Fermi energy is $-0.2$ [eV] $\ltsim E_F \ltsim$ 0.2 [eV], the proportion relation between $z$ component of the spin density and the Fermi energy is clearly seen as we expected from our analysis with the continuum model, Eq.(\ref{cit_ll}), whereas the perpendicular component to the magnetic field is always zero, independent of $E_F$.
Within this energy scale, the Landau levels shown in Fig.\ref{fig2} (b) behave as the Eq.(\ref{eq_LL}), namely the linearized continuum model where the chirality is a good quantum number describes the spectrum well.
On the other hand, when $|E_F|\gtsim0.2$ [eV], the spin density deviates from $E_F$-linear dependence.
In this regime, two Landau level branches with opposite chirality hybridize while they are approaching $k_z=0$ point.
In above argument based on the continuum model, we have neglected the hybridization of the chirality, therefore our theory, Eq.(\ref{spin_cit}), is unapplicable in this energy regime.
According to the numerical results, we conclude that the spin density response to the applied electric voltage shown in Eq. (\ref{spin_cit}) can be expected even in the lattice system when the Fermi energy is located in the linear dispersion regime.

We now discuss the relationship between the charge-induced spin torque and the chiral anomaly.
As we already mentioned in the previous section, the spin operators can be expressed in terms of the axial velocity operators in Weyl semimetals.
The correspondence suggests that we can replace the spin vertex with the axial current vertex in Fig. {\ref{FD_CIT}}, and Eq. (\ref{spin_cit}) can be interpreted as
\bea
\langle \bm j_5\rangle=\frac{e^3}{2\pi^2}\phi \bm B.
\label{j5}
\eea
According to the discussion in the previous section, Eq.(\ref{j5}) is disallowed if $\rm U(1)$ gauge symmetries are present because it explicitly depends on a gauge potential, $\phi$.
However, the presence of the chiral anomaly makes this an exceptional case.
The conservation law of axial currents can be violated even in the presence of the chiral gauge symmetry, which is known as the chiral anomaly.\cite{Adler1969,Bell1969,PeskinM.andSchroeder1995}
The violation of the axial current conservation is governed by the anomaly equation. 
By taking the divergence, we obtain
\bea
\nonumber\bm \nabla \cdot \langle\bm j_5\rangle&=&-\frac{e^3}{2\pi^2}\bm \nabla\phi \cdot \bm B=\frac{e^3}{2\pi^2}\bm E\cdot\bm B.
\label{anomaly}
\eea
This equation coincides with the anomaly equation in a static case, suggesting that Eq.(\ref{j5}) is allowed  as a part of the anomaly equation.
This indicates that the charge-induced spin torque is a consequence of the chiral anomaly.

Recently many attempts have been made to observe evidence of the chiral anomaly\cite{Nielsen1983}, for example through the observation of a negative magnetoresistance\cite{Kim2013a,Huang2015,Li2015,Xiong2015,Zhang2016,Luo2016}.
Observing a charge-induced torque is another direct piece of evidence of the chiral anomaly.
Experimentally, this phenomenon is captured as a frequency shift in the ferromagnetic resonance (FMR) peak.
In ferromagnetic metals, the interaction energy of the magnetization is given by the sum of the Zeeman energy and the exchange energy with the background conduction electron spins, 
$U=\left(-\gamma_0\rho_S SB_0+JS\langle\sigma_z\rangle\right)\hat{M}_z$ where $\rho_S$ is the density of local moments.
By substituting the contribution from the charge-induced spin torque, Eq.(\ref{spin_cit}), the interaction energy is obtained as
\bea
\nonumber U&=&-\gamma_0\rho_S S\left(1-\frac{e^2J}{2\pi^2\gamma_0\rho_S v_F}\phi\right)B_0\hat{M}_z\\
&\equiv& -\gamma_0\rho_S S\left(1-\frac{\Delta B}{B_0}\right)B_0\hat{M}_z
\eea
where $\Delta B/B_0= \frac{e^2J}{2\pi^2\gamma_0\rho_S v_F}\phi= \frac{eJ}{2\pi^2\gamma_0\rho_S v_F}\delta\mu_F$ characterizes the shift in the FMR peak.
With typical material parameters for magnetic doped topological insulators such as $\rm Cr_xBi_{2-x}Te_3$ ($J=10$ meV, $\hbar v_F=2.2$ eV$\AA^{-1}$, and $\rho_S=1.1\times 10^{-4}\AA^{-3}$ with chromium concentration $x=0.1$), we qualitatively estimate the shift in the FMR peak as
\bea
\frac{\Delta B}{B_0}\approx 0.3\times\left(\frac{\delta\mu_F}{\rm[eV]}\right).
\eea
This indicates that when a FMR measurement is performed while changing the gate voltage, the FMR peak shifts linearly in the applied gate voltage.
If the chemical potential is shifted 0.1 [eV] by the gate voltage, the shift of the FMR peak can be up to $3\%$.

\section{Discussion and conclusion}
In this paper, we have studied electrically-induced spin torques, charge-induced spin torque, spin-orbit torque, spin-transfer torque, and $\beta$-term.
They are summarized as
\bea
\bm T_e=\sigma_{\rm AHE}\phi \hat{\bm M}\times \bm B +\chi_{\rm SOT}\hat{\bm M}\times\bm E
\eea
where $\chi_{\rm SOT}=(eJSE_F\Delta)/(3\pi^2v_F\eta)$, and the spin-transfer torque and the $\beta$-term are absent because of the chiral gauge symmetry.
The first term denotes the charge-induced torque generated by an applied voltage, while the second term is the spin-orbit torque driven by an electric field due to the spin-momentum locking.
These two terms are both electrically-driven spin torques, however, the origins are completely different.
The charge-induced spin torque is an adiabatic phenomenon in which all states below Fermi energy contribute.
On the other hand, the spin-orbit torque is a non-equilibrium spin torque, associated with electrons near the Fermi surface.

In our analysis, we have taken account only of intra-nodal scattering.
This assumption is valid when the Fermi energy is located close to the Weyl nodes and the two nodes are well separated.
The absence of the spin-transfer torque and the $\beta$-terms is guaranteed as long as this assumption is valid.
When the Fermi energy is far from the Weyl nodes or the distance between nodes is small, states with different chirality hybridize and chiral gauge symmetry is no longer present.
In this case, there is no symmetry restriction, and the spin-transfer torque and the $\beta$-terms are also to be expected.

It is straight forward to introduce vertex corrections.
Several theoretical papers have concluded that, in Weyl semimetals, ladder type corrections only modify the Fermi velocity of the current vertex.\cite{Biswas2014,Burkov2015,Ominato2015}
Because our model includes the correspondence between the velocity and the spin operators, the vertex is also only modified in the coefficient.
Therefore our results are not cancelled by the vertex correction, and valid after a modification in the Fermi velocity $v_F\rightarrow \Lambda_F$, being the renormalized velocity.

Finally we discuss other spin torques within the chiral transport picture.
As we introduced in Eq.(\ref{eq_st}), the amplitude of the spin and the Gilbert damping are modified by the conducting electrons.
They are expressed as
$
\langle \bm \sigma\rangle= (\delta\alpha/JS)\delta\dot{\bm M}+(\delta S/JS)\hat{\bm z}\times \delta\dot{\bm M}
$
where $\delta \alpha$ and $\delta S$ are the modification to the Gilbert damping constant and the amplitude of spins, respectively.
Here we take the saturated magnetization axis in $z$ axis.
In the chiral transport picture, the coefficients are nothing but the longitude and Hall conductivity for the axial currents, expressed as
$
\langle \bm j_5\rangle=(ev_F/JS)a_0\bm e +(ev_F/JS)b_0\hat{\bm z}\times\bm e
$
where $\bm e=\delta\dot{\bm M}$ is the chiral electric field.
If two Weyl nodes are well separated, the coefficients are equivalent to the conductivities of the vector currents.
Namely, the correction to the Gilbert damping constant and the amplitude of spin are explicitly written as $\delta \alpha=(JS/ev_F)\sigma_{xx}$ and $\delta S=(JS/ev_F)\sigma_{xy}$.
These relations coincide with those of a ferromagnetic insulator deposited on the surface of a topological insulator.\cite{Sakai2014}

In conclusion, we microscopically derived the electrically induced spin torques in magnetically doped Weyl semimetals.
Firstly,  we examine current-induced spin torques.
Because of the spin-transport correspondence and the local chiral gauge symmetry, we found that the spin-transfer torque and the $\beta$-terms are absent as long as the two Weyl nodes with opposite chirality is well separated in momentum space.
Consequently, in the magnetic Weyl semimetals, only the spin-orbit torque is expected as current-induced spin torques up to first order with respect to spatial and temporal derivation and electrical currents.
We derived the analytical expression and found the non-equilibrium spin density is induced in the applied electric field direction when the inversion symmetry is broken, in other words, $\Delta\ne0$.
The phenomenon is understood by the Rashba-Edelstein effect because of the spin-momentum locking.

Secondly, by using a linear response theory, we derived the analytical expression of the charge-induced spin torque induced by the gate voltage in the presence an external magnetic field, and succeeded to reproduce our previous results obtained phenomenologically.
We also discussed that the charge-induced spin torque is understood in terms of the chiral anomaly and the correspondence between the spin phenomena and the chiral transport phenomena.
This indicates that the observation of the charge-induced spin torque might be direct evidence of the chiral anomaly.
Experimentally, it is expected to be observed as a frequency shift of the ferromagnetic resonance peak which is estimated as $\sim 3\%$.

\section*{Acknowledgments}
We thank Hiroshi Kohno, Yasufumi Araki, Joseph Barker, Yuya Ominato, and Takahiro Chiba for useful discussions and comments.
D.K. is supported by a JSPS Research Fellowship for Young Scientists.
This work was supported by JSPS KAKENHI Grants No. JP15H05854 and No. JP26400308.

\appendix
\begin{widetext}

\section{Current-induced spin torques}
\subsection{Spin-orbit torque}
The detail calculations for the spin-orbit torque is presented in this section.
In this section, $\mu,\nu,\rho$ and $\gamma$ are dummy indices, running over $1\sim3$ and summed in all the values of the indices.
Let us start with the Eq.(\ref{res_sot}), and proceed our calculation as
\bea
\nonumber K_\alpha^i(i\Omega_m)&=&k_BTev_F\sum_{n,\bm k,\lambda}\lambda{\rm Tr}\left[\sigma_\alpha G_\lambda\sigma_i G_\lambda^\dagger\right]\\
\nonumber&=&k_BT ev_F\sum_{n,\bm k,\lambda}\lambda D_\lambda(i\omega_n+i\Omega_m,\bm k)D_\lambda(i\omega_n,\bm k){\rm Tr}\left[\sigma_\alpha g_{0\lambda}(i\omega_n+i\Omega_m,\bm k)\sigma_i g_{0\lambda}(i\omega_n,\bm k) +v_F^2k_\mu k_\nu \sigma_\alpha \sigma _\mu \sigma_i\sigma_\nu \right]\\
&=&k_BT ev_F\sum_{n,\bm k,\lambda}\lambda D_\lambda(i\omega_n+i\Omega_m,\bm k)D_\lambda(i\omega_n,\bm k)\left\{g_{0\lambda}(i\omega_n+i\Omega_m,\bm k)g_{0\lambda}(i\omega_n,\bm k){\rm Tr}\left[\sigma_\alpha\sigma_i\right]+v_F^2\frac{k^2}{3}\delta_{\mu\nu}{\rm Tr}\left[\sigma_\alpha\sigma_\mu\sigma_i\sigma_\nu\right]\right\}
\label{b1}
\eea
after dropping terms which are odd in momentum, where $D_\lambda(i\omega_n,\bm k), g_{0\lambda}(i\omega_n,\bm k)$ are defined by 
\bea
D_\lambda(i\omega_n,\bm k)&\equiv& \left[\left(i\omega_n+E_F+\lambda \Delta-\Sigma_\lambda(i\omega_n,\bm k)\right)^2-v_F^2\bm k^2\right]^{-1}
\label{gf1}\\
g_{0\lambda}(i\omega_n,\bm k)&\equiv&i\omega_n+E_F+\lambda \Delta-\Sigma_\lambda(i\omega_n,\bm k)
\label{gf2}
\eea
satisfying
$
G_\lambda (i\omega_n,\bm k)=D_\lambda(i\omega_n,\bm k)\left(g_{0\lambda}+\lambda v_F\bm k\cdot\bm \sigma\right),
$
and we have used the relation
$
k_\mu k_\nu=\frac{k^2}{3}\delta_{\mu\nu}
$
under symmetrical integration.
The trace calculations are performed as
\bea
{\rm Tr}\left[\sigma_\alpha\sigma_i\right]=2\delta_{\alpha i}\ \ ,\ \ \delta_{\mu\nu}{\rm Tr}\left[\sigma_\alpha\sigma_\mu\sigma_i\sigma_\nu\right]=-2\delta_{\alpha i}.
\eea
By defining the Green's function projected onto conduction and valence band as
\bea
g_{\lambda,\pm}(i\omega_n)\equiv\left[i\omega_n+E_F+\lambda\Delta\mp v_F k +i\eta {\rm sgn}(\omega_n)\right]^{-1}
\eea
and using the following relations,
\bea
D_\lambda(i\omega_n,\bm k)g_{0\lambda}(i\omega_n,\bm k)&=&\frac{1}{2}\left[g_{\lambda,+}(i\omega_n)+g_{\lambda,-}(i\omega_n)\right]\\
D_\lambda(i\omega_n,\bm k)v_Fk&=&\frac{1}{2}\left[g_{\lambda,+}(i\omega_n)-g_{\lambda,-}(i\omega_n)\right],
\eea
the response function is expressed by
\bea
\nonumber K_\alpha^i(i\Omega_m)&=&k_BTev_F\delta_{\alpha i}\sum_{n,\bm k,\lambda}\lambda\left[\frac{1}{3}\left\{g_{\lambda,+}(i\omega_n)g_{\lambda,+}(i\omega_n+i\Omega_m)+g_{\lambda,-}(i\omega_n)g_{\lambda,-}(i\omega_n+i\Omega_m)\right\}\right.\\
&&\hspace{5cm}\left .+\frac{2}{3}\left\{g_{\lambda,+}(i\omega_n)g_{\lambda,-}(i\omega_n+i\Omega_m)+g_{\lambda,-}(i\omega_n)g_{\lambda,+}(i\omega_n+i\Omega_m)\right\}\right]
\label{b4}
\eea
Next we are going to perform the Matsubara summation, 
\bea
\nonumber k_BT\sum_{n}g_{\lambda,\pm}(i\omega_n)g_{\lambda,\pm}(i\omega_n+i\Omega_m)&=&\frac{i}{2\pi}\int  d\epsilon f(\epsilon)\left[\left\{g_{\lambda,\pm}^R(\epsilon)-g_{\lambda,\pm}^A(\epsilon)\right\}g_{\lambda,\pm}^R(\epsilon+\Omega)+g_{\lambda,\pm}^A(\epsilon-\Omega)\left\{g_{\lambda,\pm}^R(\epsilon)-g_{\lambda,\pm}^A(\epsilon)\right\}\right]\\
\nonumber&=&\frac{i}{2\pi}\int d\epsilon\left[\left\{f(\epsilon+\Omega)-f(\epsilon)\right\}g_{\lambda,\pm}^R(\epsilon+\Omega)g_{\lambda,\pm}^A(\epsilon)+f(\epsilon)\left\{g_{\lambda,\pm}^R(\epsilon)g_{\lambda,\pm}^R(\epsilon+\Omega)-g_{\lambda,\pm}^A(\epsilon-\Omega)g_{\lambda,\pm}^A(\epsilon)\right\} \right]\\
\nonumber&\approx&\frac{i}{2\pi}\int d\epsilon f(\epsilon)\left[(g_{\lambda,\pm}^R)^2-(g_{\lambda,\pm}^A)^2\right]+\frac{i\Omega}{2\pi} \int d\epsilon \left[f'(\epsilon)g_{\lambda,\pm}^Rg_{\lambda,\pm}^A-f(\epsilon)\left\{(g_{\lambda,\pm}^R)^3+g_{\lambda,\pm}^R)^3\right\}\right]+\mathcal{O}(\Omega^2)\\
\nonumber&=&\frac{i}{2\pi}\int d\epsilon f'(\epsilon)\left(g_{\lambda,\pm}^R-g_{\lambda,\pm}^A\right)+\frac{i\Omega}{2\pi}\int d\epsilon f'(\epsilon)\left[g_{\lambda,\pm}^Rg_{\lambda,\pm}^A-\frac{1}{2}\left\{(g_{\lambda,\pm}^R)^2+(g_{\lambda,\pm}^A)^2\right\}\right]\\
&=&-\frac{i}{2\pi}\left(g_{\lambda,\pm}^R-g_{\lambda,\pm}^A\right)-\frac{i\Omega}{4\pi}\left(g_{\lambda,\pm}^R-g_{\lambda,\pm}^A\right)^2
\label{b2}
\eea
\bea
\nonumber k_BT\sum_{n}g_{\lambda,\pm}(i\omega_n)g_{\lambda,\mp}(i\omega_n+i\Omega_m)&=&\frac{i}{2\pi}\int d\epsilon f(\epsilon)\left[\left\{g_{\lambda,\pm}^R(\epsilon)-g_{\lambda,\pm}^A(\epsilon)\right\}g_{\lambda,\mp}^R(\epsilon+\Omega)+g_{\lambda,\pm}(\epsilon-\Omega)\left\{g_{\lambda,\mp}^R(\epsilon)-g_{\lambda,\mp}^A(\epsilon)\right\}\right]\\
\nonumber&\approx&\frac{i}{2\pi}\int d\epsilon f(\epsilon)\left(g_{\lambda,\pm}^Rg_{\lambda,\mp}^R-g_{\lambda,\pm}^Ag_{\lambda,\mp}^A\right)\\
&&\hspace{1.5cm}+\frac{i\Omega}{2\pi}\int d\epsilon\left[ f'(\epsilon)g_{\lambda,\mp}^Rg_{\lambda,\pm}^A-f(\epsilon)\left\{g_{\lambda,\pm}^R(g_{\lambda,\mp}^R)^2+(g_{\lambda,\pm}^A)^2g_{\lambda,\mp}^A\right\}\right]+\mathcal{O}(\Omega^2)
\eea
\bea
\nonumber k_BT\sum_{n}\left[g_{\lambda,+}(i\omega_n)g_{\lambda,-}(i\omega_n+i\Omega_m)+g_{\lambda,-}(i\omega_n)g_{\lambda,+}(i\omega_n+i\Omega_m)\right]&=&\frac{i}{\pi}\int d\epsilon f(\epsilon)\left(g_{\lambda,+}^Rg_{\lambda,-}^R-g_{\lambda,+}^Ag_{\lambda,-}^A\right)\\
\nonumber &&+\frac{i\Omega}{2\pi}\int d\epsilon \left[ f'(\epsilon)g_{\lambda,+}^Rg_{\lambda,-}^A-f(\epsilon)g_{\lambda,+}^Rg_{\lambda,-}^R\left(g_{\lambda,+}^R+g_{\lambda,-}^R\right)+c.c\right]\\
\nonumber &=&\mathcal{O}(\Omega^0)+\frac{i\Omega}{2\pi}\int d\epsilon f'(\epsilon)\left[g_{\lambda,+}^Rg_{\lambda,-}^A-g_{\lambda,+}^Rg_{\lambda,-}^R+c.c\right]\\
\nonumber&=&\mathcal{O}(\Omega^0)-\frac{i\Omega}{2\pi}\int d\epsilon f'(\epsilon)\left(g_{\lambda,+}^R-g_{\lambda,+}^A\right)\left(g_{\lambda,-}^R-g_{\lambda,-}^A\right)\\
&=&\mathcal{O}(\Omega^0)+\frac{i\Omega}{2\pi}\left(g_{\lambda,+}^R-g_{\lambda,+}^A\right)\left(g_{\lambda,-}^R-g_{\lambda,-}^A\right)
\label{b3}
\eea
where $g_{\lambda,\pm}^{R,(A)}\equiv g_{\lambda,\pm}^{R,(A)}(\epsilon)$, $f(\epsilon)$ is the Fermi-Dirac distribution function, and we have substituted $i\Omega_m\rightarrow \Omega$.
We have also used the relation $\frac{d}{d\epsilon}g_{\lambda,\pm}^{R,(A)}=-(g_{\lambda,\pm}^{R,(A)})^2$, and assumed zero temperature, namely $f(\epsilon)=-\delta(\epsilon)$.
In the weak scattering regime where $\eta<<E_F$, the Green's functions can be expanded by $g_{\lambda,\pm}^R-g_{\lambda,\pm}^A=-2\pi i\delta(E_F+\lambda\Delta\mp v_F k)$, and momentum integrals are evaluated as
\bea
\nonumber \sum_{\bm k,\lambda}({\rm \ref{b2}})&=&\sum_\lambda \lambda\int \frac{d k^3}{(2\pi)^3}\frac{i\Omega}{4\pi}\left[\left(g_{\lambda,+}^R-g_{\lambda,+}^A\right)^2+\left(g_{\lambda,-}^R-g_{\lambda,-}^A\right)^2\right]\\
\nonumber &\approx&-\frac{\Omega}{4\pi^2 v_F^3}\sum_\lambda  \lambda\int d\xi \xi^2\left[\delta(E_F+\lambda\Delta- \xi)\left(\frac{1}{E_F+\lambda\Delta-\xi+i\eta}-\frac{1}{E_F+\lambda\Delta-\xi-i\eta}\right)\right.\\
\nonumber &&\hspace{4cm}\left.+\delta(E_F+\lambda\Delta+ \xi)\left(\frac{1}{E_F+\lambda\Delta+\xi+i\eta}-\frac{1}{E_F+\lambda\Delta+\xi-i\eta}\right)\right]\\
&=&\frac{i\Omega}{2\pi^2 v_F^3}\sum_{\lambda}  \lambda\frac{\left(E_F+\lambda\Delta\right)}{\eta}=i\Omega\frac{E_F\Delta}{\pi^2 v_F^3\eta},\\
\nonumber \sum_{\bm k,\lambda}({\rm \ref{b3}})&=&\sum_\lambda  \lambda\int \frac{d k^3}{(2\pi)^3}\frac{i\Omega}{2\pi}\left(g_{\lambda,+}^R-g_{\lambda,+}^A\right)\left(g_{\lambda,-}^R-g_{\lambda,-}^A\right)\\
\nonumber&\approx&\frac{\Omega}{2\pi^2v_F^3}\sum_\lambda \lambda \int d\xi \xi^2\frac{i}{2\pi}\left(-2\pi i\delta(E_F+\lambda \Delta-\xi)-\frac{i\eta}{(E_F+\lambda \Delta-\xi)^2+\eta^2}\right)\\
\nonumber&&\hspace{4cm}\times\left(-2\pi i\delta(E_F+\lambda \Delta+\xi)-\frac{i\eta}{(E_F+\lambda \Delta+\xi)^2+\eta^2}\right)\\
\nonumber &=&-\frac{i\Omega}{2\pi^2v_F^3}\sum_\lambda \lambda\int d\xi \xi^2\left[\delta(E_F+\lambda \Delta-\xi)\frac{\eta}{(E_F+\lambda \Delta+\xi)^2}+\delta(E_F+\lambda \Delta+\xi)\frac{\eta}{(E_F+\lambda \Delta-\xi)^2}\right]\\
&=&-\frac{i\Omega}{2\pi^2v_F^3}\sum_\lambda \lambda\frac{\eta}{4}=0
\eea
where we have neglected $\mathcal{O}(\Omega^0)$ terms because they are cancelled by the definition of the spin susceptibility, Eq.(\ref{spin_current}).
By substituting these result for Eq.(\ref{b4}), the response function is obtained by
\bea
K_\alpha^i(\Omega)=i\Omega\frac{eE_F\Delta}{3\pi^2 v_F^2\eta}
\eea
which is presented in Eq.(\ref{k_sot}).

\subsection{Spin-transfer torque and $\beta$-term}
In this section, we present the detail calculations for obtaining Eq.(\ref{k_current}).
Let us start with the first term of Eq.(\ref{res_current}),
\bea
\nonumber K_{\alpha \beta}^{ij}(i\Omega_m)&=&k_BTeJSv_F^2\sum_{n,\bm k,\lambda }{\rm Tr}\left[\sigma_\alpha G_\lambda \sigma_jG_\lambda\sigma_\beta G_\lambda \sigma_i  G^+_\lambda\right]\\
\nonumber &=&k_BTeJSv_F^2\sum_{n,\bm k,\lambda}D_{\lambda}(i\omega_n+i\Omega_m,\bm k)D_{\lambda}(i\omega_n,\bm k)^3\left\{g_{0\lambda}(i\omega_n,\bm k)^3g_{0\lambda}(i\omega_n+\Omega_m,\bm k){\rm Tr}\left[\sigma_\alpha\sigma_j\sigma_\beta\sigma_i\right]\right.\\
\nonumber&&+\delta_{\mu\nu}\frac{(v_Fk)^2}{3}g_{0\lambda}(i\omega_n,\bm k)^2\left({\rm Tr}\left[\sigma_\alpha\sigma_j\sigma_\beta\sigma_\mu\sigma_i\sigma_\nu\right]+{\rm Tr}\left[\sigma_\alpha\sigma_j\sigma_\mu\sigma_\beta\sigma_i\sigma_\nu\right]+{\rm Tr}\left[\sigma_\alpha\sigma_\mu\sigma_j\sigma_\beta\sigma_i\sigma_\nu\right]\right)\\
\nonumber&&+\delta_{\mu\nu}\frac{(v_Fk)^2}{3}g_{0\lambda}(i\omega_n,\bm k)g_{0\lambda}(i\omega_n+i\Omega_m,\bm k)\left({\rm Tr}\left[\sigma_\alpha\sigma_j\sigma_\mu\sigma_\beta\sigma_\nu\sigma_i\right]+{\rm Tr}\left[\sigma_\alpha\sigma_\mu\sigma_j\sigma_\beta\sigma_\nu\sigma_i\right]+{\rm Tr}\left[\sigma_\alpha\sigma_\mu\sigma_j\sigma_\nu\sigma_\beta\sigma_i\right]
\right)\\
\nonumber &&\left. +v_F^4\frac{k^2k^2}{15}\left(\delta_{\mu\nu}\delta_{\rho\gamma}+\delta_{\mu\rho}\delta_{\nu\gamma}+\delta_{\mu\gamma}\delta_{\nu\rho}\right){\rm Tr}\left[\sigma_\alpha\sigma_\mu\sigma_j\sigma_\nu\sigma_\beta\sigma_\rho\sigma_i\sigma_\gamma\right]\right\}.
\eea
The second term can be estimated by just exchanging $i\leftrightarrow j$ and replacing $\Omega_m\rightarrow -\Omega_m$.
The equation looks cumbersome, however the Pauli matrices always appear with even number in trace.
This constrains the response function to being proportional to pairs of Kronecker deltas with all possible combination of indices,
\bea
K_{\alpha \beta}^{ij}(i\Omega_m)=A\delta_{\alpha\beta}\delta_{ij}+B\delta_{\alpha i}\delta_{\beta j}+C\delta_{\alpha j}\delta_{\beta i}
\eea
where $A,B,C$ are constants of proportionality.

\section{Charge-induced torque}
The detail calculations for the charge-induced spin torque is presented in here.
Because the charge-induced spin torque is an adiabatic spin torque, namely the leading order of the spin torque is give in $\mathcal{O}(\eta^0)$.
We focus on the non dissipation case, $\eta=0$.
We also assume that the Fermi energy is located on the Weyl nodes, $E_F=0$.
For calculating the charge-induced spin torque, it is easier to introduce the Dirac's Gamma matrices and express the Green's function as
$
G(k)=[v_Fk_i\gamma_i]^{-1}
$
where $i=1\sim4$, $k=(\bm k,\omega/v_F)$ is a four-momentum and the gamma matrices are defined as
\bea
\gamma_i\equiv i\tau_y\sigma_i\ (i=1,2,3)\ ,\ \gamma_4\equiv-i\tau_x
\eea
satisfying $\{\gamma_i,\gamma_j\}=2\eta_{ij}$ with the Euclidean metric $\eta_{ij}={\rm diag}(-1,-1,-1,-1)$.\cite{Zyuzin2012a}
The interaction with the electromagnetic fields and the magnetization are given as
\bea
H_{\rm EM}&=&-\int d^3r\ ev_F A_i\overline{\psi}i\gamma_i\psi\\
H_{\rm exc}&=&-\int d^3r\ JS\hat{\bm M}\cdot \overline{\psi}i\bm \gamma\gamma_5\psi
\eea
where $A_i=(\bm A,\phi/v_F)$ for $i=1\sim4$, and $\overline{\psi}\equiv \psi^\dagger(\bm r) \tau_x,\ \psi\equiv c(\bm r)$, and $\gamma_5\equiv\gamma_1\gamma_2\gamma_3\gamma_4= \tau_z$, thus the current and spin operators are given by
\bea
\nonumber j_i=\frac{\delta H_{\rm EM}}{\delta A_i}=-iev_F\overline{\psi}i\gamma_i\psi,\ \sigma_i=\frac{\delta H_{\rm exc}}{\delta (JSM_i)}=-i\overline{\psi}\gamma_i\gamma_5\psi.
\eea
With this notation, Eq.(\ref{cit_chi}) is rewritten as
\bea
\chi_\alpha^i(\bm q)&=&i\frac{e^2}{v_F}\int \frac{d^4 k}{(2\pi^4)}{\rm Tr}\left[\gamma_\alpha\gamma_5\frac{\slash{k}}{k^2}\gamma_4\frac{\slash{k}}{k^2}\gamma_i\frac{(\slash{k}+\slash{q})}{(k+q)^2}\right.
 \left. +\gamma_\alpha\gamma_5\frac{(\slash{k}-\slash{q})}{(k-q)^2}\gamma_i\frac{\slash{k}}{k^2}\gamma_4\frac{\slash{k}}{k^2}\right]
\label{a1}
\eea
where four-momenta are expressed in the Feynman's slash notation, $\slash{k}=k_i\gamma_i$.
Here the $\alpha=1\sim3$ and $i=1\sim4$ denote the spin and velocity direction respectively.
Note that, because the integral in Eq.(\ref{a1}) is divergent, it should be evaluated with a proper regularization scheme.
In this paper, we introduce dimensional regularization.
The basic idea of the dimensional regularization is to perform an integral of an analytic function in the space-time dimension $d$ where the integral converges, then taking limit $d\rightarrow 4$ after obtaining the analytical expression in $d$-dimension.
By expanding Eq.(\ref{a1}) in terms of the incident momentum $\bm q$, one can obtain
$
\chi_\alpha^i(\bm q)=i(e^2/v_F)\left(\Pi_{\alpha i}-\Pi_{\alpha i j}q_j\right)+\mathcal{O}(q^2)
$
where
\bea
\Pi_{\alpha i}&=&\int \frac{d^dk}{(2\pi)^d}\frac{k_\mu k_\nu k_\rho}{k^6}{\rm Tr}\left[\gamma_\alpha\gamma_5\gamma_\mu \gamma_4\gamma_\nu \gamma_i\gamma_\rho \right]+(i\leftrightarrow 4)
\label{a2}\\
\nonumber \Pi_{\alpha i j}&=&\int\frac{d^dk}{(2\pi)^d} \frac{k_\mu k_\nu k_\rho k_\lambda}{k^8}{\rm Tr}\left[\gamma_\alpha\gamma_5\gamma_\mu \gamma_4\gamma_\nu \gamma_i\gamma_\rho \gamma_j\gamma_\lambda \right]-(j\leftrightarrow 4)\\\label{a3}
\eea
with dummy indices $\mu,\nu,\rho,\lambda=1\sim d$.
Since the integral is odd in four-momentum, $\mathcal{O}(q^0)$ term is zero, $\Pi_{\alpha i}=0$.
$\mathcal{O}(q^1)$ term, therefore, becomes the leading order in incident momentum, which is evaluated as
\bea
\Pi_{\alpha i j}&=&\frac{1}{d(d+2)}\int\frac{d^dk}{(2\pi)^d} \frac{1}{k^4}\left(\mathcal{T}_{\alpha 4 i j}-\mathcal{T}_{\alpha j i 4}\right)
\label{a5}\\
\nonumber \mathcal{T}_{ijkl}&\equiv&-\left(\delta_{\mu\nu}\delta_{\rho\lambda}+\delta_{\mu\rho}\delta_{\nu\lambda}+\delta_{\nu\rho}\delta_{\mu\lambda}\right){\rm Tr}\left[\gamma_5\gamma_i\gamma_\mu \gamma_j\gamma_\nu \gamma_k\gamma_\rho \gamma_l\gamma_\lambda \right] 
\eea
where we have used
\bea
k_\mu k_\nu k_\rho k_\lambda=\frac{k^4}{d(d+2)}\left(\delta_{\mu\nu}\delta_{\rho\lambda}+\delta_{\mu\rho}\delta_{\nu\lambda}+\delta_{\nu\rho}\delta_{\mu\lambda}\right)
\eea
under the symmetrical integration.
To evaluate the trace, we have to take care of the commutation relation between $\gamma_5$ and $\gamma_\mu$ because $\gamma_5$ is essentially defined in four-dimension.
If we impose all gamma matrices to satisfy the anticommutation relations, $\gamma_5$ commutes with $\gamma_\mu$ for $\mu=1\sim 4$ while commutes with $\gamma_\mu$ for other $\mu$.
Keeping this in mind, trace parts with Kronecker delta are evaluated as
\bea
\nonumber \mathcal{T}_{\alpha 4 i j}&=&(d-4)(d+2){\rm Tr}\left[\gamma_5\gamma_4\gamma_\alpha \gamma_i\gamma_j\right]\\
&\overset{d\rightarrow 4}{=}&-4(d-4)(d+2)\epsilon_{\alpha ij}
\label{a6}
\eea
The momentum integral in $d$-dimension is also performed as
\bea
\nonumber \int\frac{d^dk}{(2\pi)^d} \frac{1}{k^4}&=&-\frac{1}{(4\pi)^{d/2}}\frac{\Gamma(2-\frac{d}{2})}{\Gamma(2)}\\
&\overset{d\rightarrow 4}{=}&\frac{1}{(4\pi)^2}\left. \left(\frac{1}{d/2-2}-\gamma\right)\right|_{d\rightarrow4}
\label{a7}
\eea
where we have used the expansion of the Gamma function near the pole, $\Gamma(x)\approx 1/x-\gamma$, and $\gamma\approx0.5772$ is the Euler-Mascheroni constant.
By substituting Eq.(\ref{a6}) and (\ref{a7}) for Eq. (\ref{a5}), we finally obtain
\bea
\nonumber \Pi_{\alpha i j}&=&\left .-\frac{16}{(4\pi^2)}\epsilon_{\alpha i j}\frac{1}{d(d+2)}(d-4)(d+2)\frac{1}{d-4}\right|_{d\rightarrow 4}\\
&=&-\frac{1}{4\pi^2}\epsilon_{\alpha i j},
\eea
and, as a result, the susceptibility, Eq. (\ref{cit_chi2}), is obtained.

\end{widetext}
%


\begin{thebibliography}{56}
\bibitem{note1}
The Hamiltonian, Eq.(\ref{H_0}), does not have the local chiral gauge symmetry. However, it can be mapped to the locally chiral gauge symmetric Hamiltonian of Weyl electrons coupled to the axial vector potentials, when the inter-nodal scattering is neglected which is the case in our analysis.
\end{thebibliography}
\end{document}